# Catalytic growth of ultralong graphene nanoribbons on insulating substrates

Bosai Lyu[1,2†], Jiajun Chen[1,2†], Shuo Lou[1,2†], Can Li[1,2†], Lu Qiu[3,4], Wengen Ouyang[5†], Jingxu Xie[1,2], Izaac Mitchell[3,4], Tongyao Wu[1,2], Aolin Deng[1,2], Cheng Hu[1,2], Xianliang Zhou[1,2], Peiyue Shen[1,2], Saiqun Ma[1,2], Zhenghan Wu[1,2], Kenji Watanabe[6], Takashi Taniguchi[7], Xiaoqun Wang[1,2,8], Qi Liang[1,2,8], Jinfeng Jia[1,2,8], Michael Urbakh[9], Oded Hod[9*], Feng Ding[3,4*], Shiyong Wang[1,2,8*], Zhiwen Shi[1,2,8*]

[1]Key Laboratory of Artificial Structures and Quantum Control (Ministry of Education), Shenyang National Laboratory for Materials Science, School of Physics and Astronomy, Shanghai Jiao Tong University, Shanghai 200240, China.
[2]Collaborative Innovation Centre of Advanced Microstructures, Nanjing University, Nanjing 210093, China.
[3]Centre for Multidimensional Carbon Materials, Institute for Basic Science, Ulsan 44919, South Korea.
[4]School of Materials Science and Engineering, Ulsan National Institute of Science and Technology, Ulsan 44919, South Korea
[5]Department of Engineering Mechanics, School of Civil Engineering, Wuhan University, Wuhan, Hubei 430072, China.
[6]Research Centre for Functional Materials, National Institute for Materials Science, 1-1 Namiki, Tsukuba 305-0044, Japan.
[7]International Centre for Materials Nanoarchitectonics, National Institute for Materials Science, 1-1 Namiki, Tsukuba 305-0044, Japan.
[8]Tsung-Dao Lee Institute, Shanghai Jiao Tong University, Shanghai, 200240, China.
[9]Department of Physical Chemistry, School of Chemistry and The Sackler Centre for Computational Molecular and Materials Science, The Raymond and Beverly Sackler Faculty of Exact Sciences, Tel Aviv University, Tel Aviv 6997801, Israel.

†These authors contributed equally to this work.
*Correspondence to: odedhod@tauex.tau.ac.il, f.ding@unist.ac.kr, shiyong.wang@sjtu.edu.cn, zwshi@sjtu.edu.cn

**Graphene nanoribbons (GNRs) with widths of a few nanometres are promising candidates for future nano-electronic applications due to their structurally tunable bandgaps, ultrahigh carrier mobilities, and exceptional stability. However, the direct growth of micrometre-long GNRs on insulating substrates, which is essential for the fabrication of nano-electronic devices, remains an immense challenge. Here, we report the epitaxial growth of GNRs on an insulating hexagonal boron nitride (*h*-BN) substrate through nanoparticle-catalysed chemical vapor deposition (CVD). Ultra-narrow GNRs with lengths of up to 10 μm are synthesized. Remarkably, the as-grown GNRs are crystallographically aligned with the *h*-BN substrate, forming one-dimensional (1D) moiré superlattices. Scanning tunnelling microscopy reveals an average width of 2 nm and a typical bandgap of ~1 eV for similar GNRs grown on conducting graphite substrates. Fully atomistic computational simulations support the experimental results and reveal a competition between the formation of GNRs and carbon nanotubes (CNTs) during the nucleation stage, and van der Waals sliding of the GNRs on the *h*-BN substrate throughout the growth stage. Our study provides a scalable, single-step method for growing micrometre-long narrow GNRs on insulating substrates, thus opening a route to explore the performance of high-quality GNR devices and the fundamental physics of 1D moiré superlattices.**

Graphene nanoribbons (GNRs) have been investigated for over two decades without exhausting their wonders and challenges, since the first theoretical prediction of their unique quantum confinement and edge physics[1]. In contrast to semimetalic bulk graphene, GNRs typically possess a finite bandgap of size depending on their edge type and width[1, 2, 3, 4]. Specifically, GNRs with armchair-type edge geometry (AC-GNRs) are predicted to have alternating bandgaps with an envelope inversely proportional to their width[2, 3], whereas zigzag-edged GNRs (ZZ-GNRs) are predicted to have exotic electronic and magnetic properties, such as spin-polarized electronic edge states[1] and half-metallicity[4, 5], thus exhibiting attractive potential in future nanoelectronic, spintronic, and quantum information technologies[6].

In light of this, great efforts have been devoted into experimentally realizing GNRs, leading to various fabrication techniques including: top-down cutting of graphene sheets into GNRs[7, 8], unzipping of carbon nanotubes (CNTs),[9] and bottom-up CVD[10-12] and on-surface synthesis of a variety of GNR structures[13, 14]. These pioneering advances allowed for the exploration of many unique attributes of GNRs[7, 10, 14]. However, the direct synthesis of micrometre-long high-quality narrow GNRs on insulating substrates, which is crucial for their utilization in nanoscale electronic and spintronic devices, remains a major challenge[15].

Here, we present a scalable single-step synthesis of micrometre-long GNRs on insulating atomically flat hexagonal boron nitride (*h*-BN) surfaces via catalytic chemical vapor deposition (CVD, see *Methods* section for details). Our technique involves nanoparticle centres (see Fig. 1a-c) that have proven capability to catalyse the growth of a variety of one-dimensional (1D) materials[16, 17]. In particular, transition metal nanoparticles, e.g. Fe, Co, and Ni, that can dissolve carbon atoms, have been extensively used for growing CNTs of length up to centimeters[18-20] with exceptional quality in terms of diameter uniformity[17, 19] and low defect density[17, 18, 21]. Similarly, our approach yields ultralong (up to 10 μm) narrow (down to 1.4 nm) GNRs with hydrogen terminated regular edges (Fig. 1e). In contrast to commonly used substrates, such as $SiO_2$/Si, quartz or sapphire, we choose atomically flat *h*-BN as the growth substrate. The *h*-BN substrate can reduce the GNR stacking energy by ~50 meV per carbon atom and simultaneously yield superlubric interfaces, thus favouring the growth of GNRs over CNTs (see Fig. 3c-d). The chemically inert flat surface and large electronic bandgap of the *h*-BN substrate preserve the ultrahigh carrier mobility of the overlying GNRs. Moreover, the grown GNRs are found to be crystallographically aligned with the *h*-BN substrate, showing well-defined 1D moiré superlattices. Hence, the obtained GNRs serve as excellent model systems to study and verify theoretical predictions of the intricate dependence of their electronic properties on their symmetry and width[1, 2, 3] as well as rich low-dimensional moiré physics.

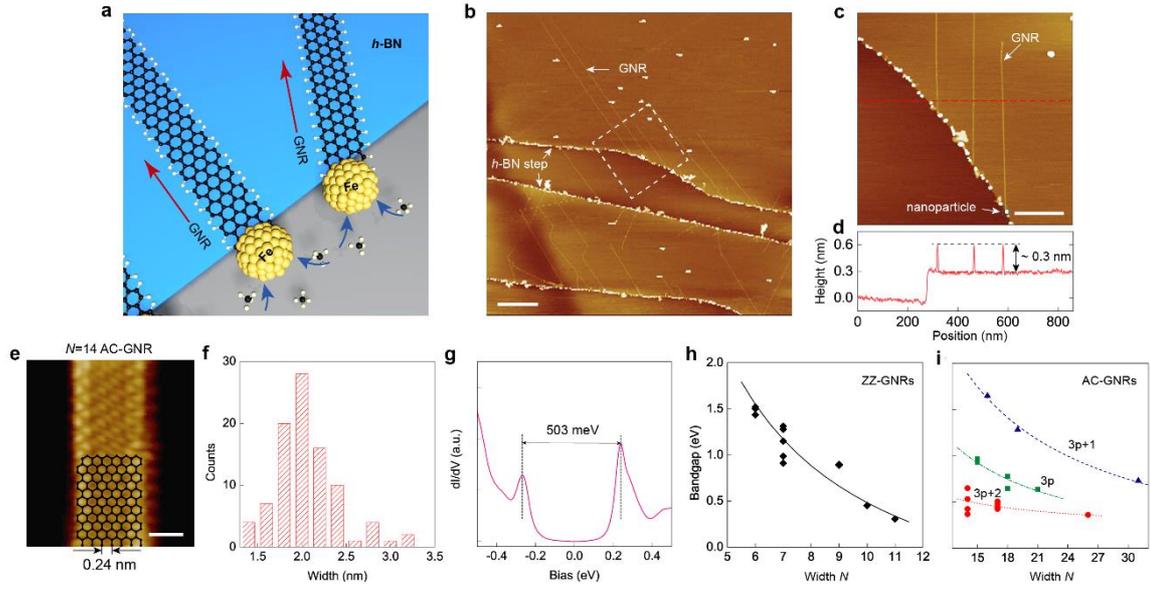

**Figure 1 Growth and characterization of high-quality graphene nanoribbons (GNRs). a,** Schematic of GNR growth on an *h*-BN substrate. The growth is catalysed by Fe nanoparticles attached to the *h*-BN step edge. **b,** A large-scale AFM topography image of as-grown GNRs (showing as bright straight lines) on *h*-BN. Scale bar: 0.5 µm. **c,** Zoom-in on the region marked by the dashed line in panel (b), rotated by 45°. Three straight micrometer-long GNRs are grown from nanoparticles attached to an *h*-BN step edge. Scale bar: 200 nm. **d,** Height profile along the dashed red line appearing in panel (c), showing that all three GNRs have a uniform height of 0.3 nm. **e,** High-resolution STM image of GNRs grown on graphite (constant current image, scanning parameters: V = 1.5 V, I = 50 pA) with partly overlaid hexagonal lattice structure of a *N*=14 armchair ribbon (width of 2 nm). Scale bar: 1 nm. **f,** Distribution histogram of the width of GNRs grown atop a graphite substrate, measured using STM. **g,** dI/dV spectroscopy of a ~0.5 eV bandgap GNR grown atop graphite. **h, i,** The bandgap measured by dI/dV spectroscopy on graphite substrate, as a function of (h) ZZ-GNR and (i) AC-GNRs width. The points represent experimental values, and the lines are ~1/*w* fittings following previous theories. For the AC-GNRs (panel i), a gap size order $\Delta_{3p+1} > \Delta_{3p} > \Delta_{3p+2}$ (p=1, 2, 3, …) can be observed.

The GNR growth is illustrated in Fig. 1a. Catalytic Fe nanoparticles are first deposited onto *h*-BN flakes on $SiO_2$/Si substrates, which are then placed in a CVD tube furnace and heated to a growth temperature of ~800 °C under a flowing hydrogen and argon mixture at atmospheric pressure. At the growth temperature, methane is injected into the furnace as a carbon source for the GNR growth. Under these growth conditions the structure of catalyst is expected to be either $Fe_3C$ or $Fe_2C$[22]. The growth typically lasts 30 minutes, followed by cooling under a protective atmosphere of hydrogen and argon. Further details regarding the growth process can be found in the *Methods* section.

The as-grown samples are initially characterized by atomic force microscopy (AFM). A typical large-scale AFM image appears in Fig. 1b demonstrating the appearance of many micrometre-long bright lines. A zoom-in on few such lines (Fig. 1c) displays that they originate from nanoparticles located at an *h*-BN edge, and are straight and possess a uniform height of ~0.3 nm (Fig. 1d). Notably, the height is smaller than the diameter of the narrowest carbon nanotube ever reported[23] and is close to the thickness of monolayer graphene (~0.34 nm), indicating that we observe the growth of GNRs. To further support this conclusion, a combination of scanning tunneling microscopy (STM) and Raman characterization was performed (Fig. S1). For the STM measurements, we directly grew GNRs on conductive graphite substrates using the same recipe as for the hBN substrates (See *Methods*). The micrometre-long GNRs have minute width variations as shown in large-scale STM images (see, e.g., Fig. S2). A high-resolution STM image of one of the bright lines shows a honeycomb lattice structure with a lattice constant of 2.46 Å (Fig. 1e), perfectly matching that of graphene, with uniform width and smooth edges. The Raman spectra of the as-grown samples (Fig. S1b), show a prominent single G-peak located at 1598 cm$^{-1}$, representing sp$^2$ carbon hybridization. Hence, both the real-space imaging and the spectral evidence demonstrate unambiguously that the as-grown samples consist of high-quality GNRs. Our STM measurements also provide information regarding the GNR width, from which we extracted the width distribution histogram appearing in Fig. 1f. We find that most grown ribbons are narrower than 3.5 nm with an average width of ~2 nm which could be a result of limitation of the nanoparticle size (Fig. S3).

Following the structural characterization, we performed scanning tunneling spectroscopy (STS) measurements to probe the GNRs' electronic bandgap. Fig. 1g shows a typical dI/dV spectrum of an AC-GNR on graphite, from which a bandgap of ~500 meV is extracted. The bandgaps of a series of ZZ- and AC-GNRs of different widths (classified by *N* - the number of dimer lines for AC-GNRs and zigzag chains for ZZ-GNRs[1]) are systematically measured (Fig. S4 and S5). For ZZ-GNRs, the bandgap is found to be inversely proportional to the ribbon width (Fig. 1h). For AC-GNRs, on top of this inverse proportionality, the bandgap shows the predicted 3-fold oscillations (Fig. 1i), by which the GNRs can be categorized into three types with different gap sizes: $\Delta_{3p+1} > \Delta_{3p} > \Delta_{3p+2}$ (p=1, 2, 3, …)[1, 2]. This further indicates the well-defined edge structures of the grown GNRs. We note that the measured GNR bandgaps in the range of 0.3-1.5 eV, are comparable with those of germanium (~0.67 eV) and silicon (~1.1 eV), and therefore are suitable for field-effect devices.

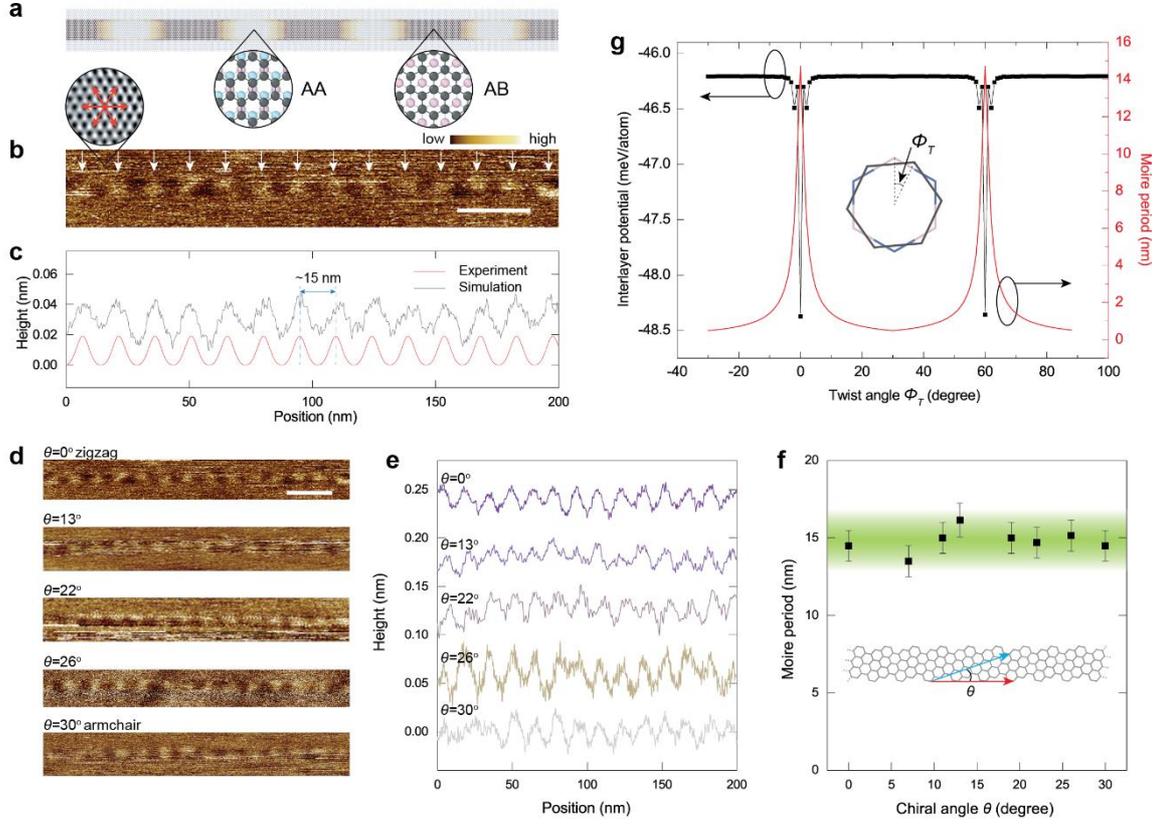

**Figure 2 Observation of 1D moiré superlattices of GNRs on *h*-BN. a,** Schematic of the 1D moiré pattern along a zigzag GNR aligned atop *h*-BN, where the colours along the GNR reflect the local height. **b,** A fine AFM scan of a zigzag GNR on an *h*-BN substrate reveals a clear 1D periodic pattern along the GNR. Scale bar: 30 nm. The inset shows the crystal orientation of the *h*-BN substrate, where the red arrows denote the zigzag directions. **c,** Experimental height profile (red) extracted from the 1D moiré superlattice in (b), showing a period of ~15 nm and a corrugation of ~0.02 nm, the same as those of aligned 2D graphene on *h*-BN. The grey line is the MD simulated height profile of the 1D moiré pattern generated when a zigzag GNR is placed on an *h*-BN surface. **d,** High-resolution AFM images of GNRs along different *h*-BN crystal orientations: $\theta = 0°$ (zigzag), 13°, 22°, 26°, and 30° (armchair), all displaying a uniform moiré pattern of ~15 nm, indicating that they align with the *h*-BN lattice. **e,** Height profiles of the GNRs presented in panel (d). **f,** A summary of moiré periods as a function of chiral angle $\theta$. **g,** The dependence of the GNR/*h*-BN interlayer stacking energy (black curve) and the moiré period (red) on the twist angle $\Phi_T$. At zero twist angle the stacking energy is minimal and the moiré period is ~15 nm, consistent with experimental observations.

Similar to the case of perfectly aligned graphene on *h*-BN[24], GNRs aligned along the zigzag direction of the *h*-BN substrate present a well-defined 1D moiré pattern with a period of ~15 nm (see Figs. 2b, c), induced by their ~1.8% lattice constant mismatch. The height corrugation along these moiré patterns is ~0.02 nm, with minima corresponding to the optimal AB-stacking and maxima to the AA-stacking mode, as illustrated in Fig. 2a. Fig. 2d shows high-resolution friction mode AFM images of GNRs along different crystal orientations of the *h*-BN substrate: $\theta = 0°$ (zigzag), 13°, 22°, 26°, and 30° (armchair) from top to bottom. Notably, regardless of the GNR orientation, pronounced 1D moiré patterns are observed (Fig. 2e) with similar periods of ~15 nm (see Fig. 2f). This indicates that all the as-grown GNRs are crystallographically well-aligned with the underlying *h*-BN substrate (red curve in Fig. 2g and Fig. S6a). Previous studies of two-dimensional (2D) moiré superlattices resulted in interesting emergent phenomena[25, 26, 27, 28, 29], such as satellite Dirac points[25], Hofstadter's butterfly[26], Mott insulators[28] and superconductivity[29]. Furthermore, 2D moiré patterns have also been observed in wide GNRs[10, 12]. The 1D graphene moiré superlattice revealed herein, with well-defined edge structure, width, and length, open the way to study unexplored one-dimensional moiré physics that is expected to present rich new phenomena.

The perfect alignment along different *h*-BN crystal directions, indicated by the uniform moiré periods, implies that the as-grown GNRs have different edge chirality, namely, GNRs growing along the zigzag (armchair) *h*-BN direction are of zigzag (armchair) type, whereas GNRs growing along other crystal directions are chiral. This is similar to the case of CNTs growing on *h*-BN or graphite substrates[21, 30]. Thus, the chirality of the GNRs can be determined by measuring their orientation relative to the *h*-BN substrate crystal (Fig. S6d). The observed perfect alignment can be rationalized by considering the fact that the stacking energy of GNRs atop *h*-BN at 0° is ~2 meV/atom lower relative to other twist angles (Fig. 2g). Such fingerprints of epitaxial growth indicate that the *h*-BN substrate plays a crucial role in the GNR growth process.

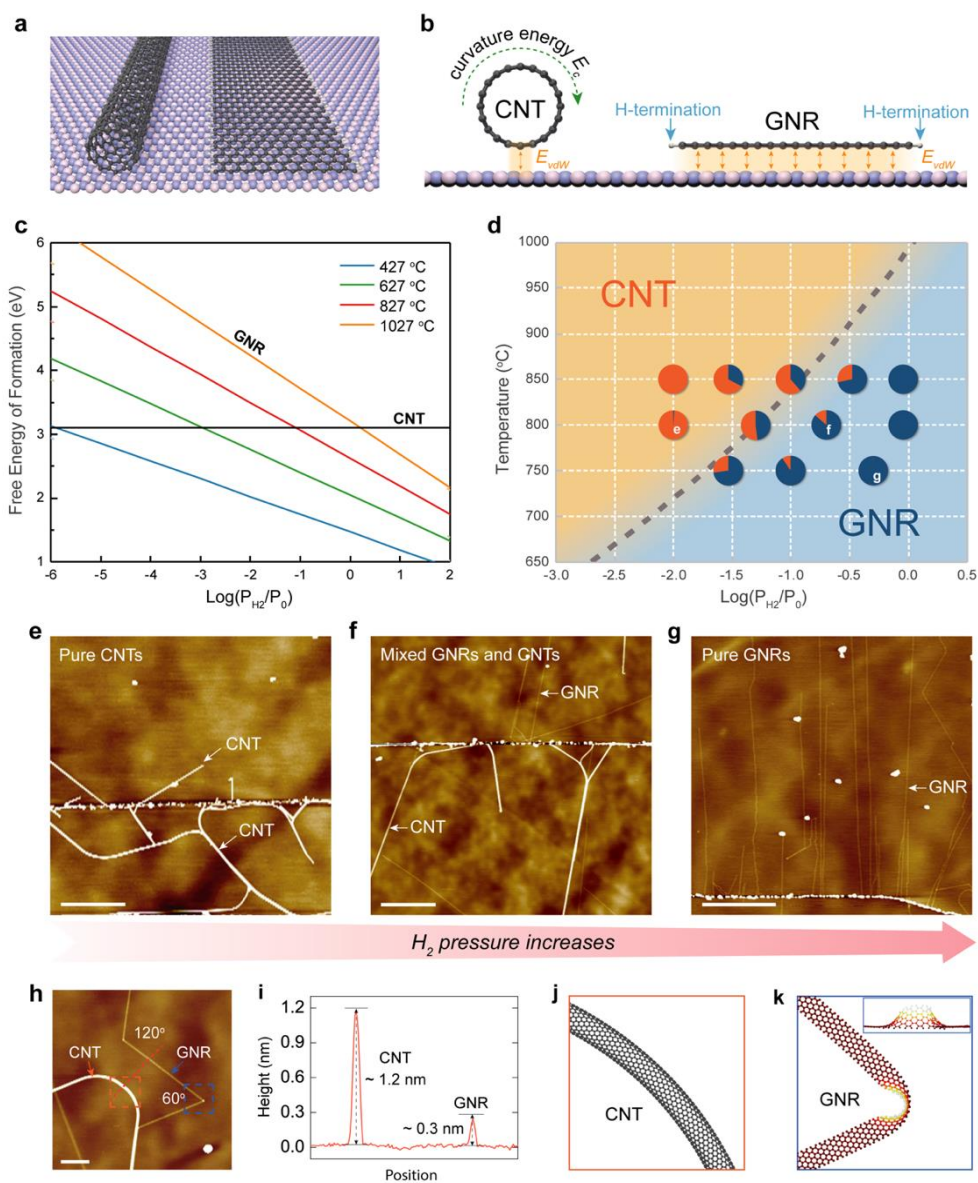

**Figure 3 Competing nucleation between GNRs and carbon nanotubes (CNTs). a, b,** Schematics of a tubular CNT and a planar GNR atop an *h*-BN substrate. **c,** The Gibbs free energies of formation of CNTs and GNRs at different temperatures and $H_2$ pressures. **d,** Theoretical growth phase diagram, where yellow and blue regions favour the growth of CNTs and GNRs, respectively. The grey dashed line represents the conditions, under which the probability of yielding GNRs and CNTs is the same. Pie charts present experimental distributions of GNR (blue) and CNT (orange) yields measured under different experimental conditions. **e-g,** AFM images of three typical growth results for different conditions as marked in (d). Scale bar, 500 nm. With the continuous increase of hydrogen pressure, a systematic increase of GNR yield is clearly observed. **h,** Close-up topography of bended GNR and CNT. The GNR turns abruptly by 60° and 120° with sharp turning corners as illustrated by top and side views in (k). The CNT turns gradually as sketched in (j). Scale bar, 100 nm. **i,** Height profile taken along the red dashed line in (h), showing a large height difference between the CNT (~1.2 nm) and the GNR (~0.3 nm).

To explore this idea, we first investigated the nucleation stage, where a competition between the growth of planar GNRs and cylindrical CNTs, was revealed (see Fig. 3a). In both cases, graphitic islands are initially formed on the surface of the catalyst nanoparticle. To produce CNTs, the graphitic islands should merge to form a cap that eventually lifts-off from the particle. This process is energetically unfavourable (compared to GNR nucleation), yet achievable when the catalytic nanoparticles are supersaturated by carbon at sufficiently high temperatures (typically > 800°C)[17-19, 31, 32]. Introducing an *h*-BN substrate enables an additional growth channel, where instead of lifting off from the nanoparticle, the graphitic island stacks atop the *h*-BN substrate and slides on its surface during the growth process, yielding planar GNRs. In favour of CNT growth is the fact that they lack elongated edges. On the contrary, narrow CNTs have a significant curvature energy contribution and a lower adhesion energy with the underlying *h*-BN substrate (due to their reduced contact area, see Fig. 3a-b), compared to that of the corresponding GNRs. The balance between all these factors determines the thermodynamically favourable nucleation path.

A quantitative analysis of the two nucleation routes requires the evaluation of the free energies of formation of CNTs and GNRs under the same thermodynamic conditions. This was achieved by considering the stacking energy, the curvature energy, and the edge hydrogen (per our CVD conditions) passivation energy at different temperatures (*T*) and under different H$_2$ gas pressures ($P_{H_2}$, see computational details in the *Methods* section and in SI section 6a). Figure 3c, compares the CNT and GNR free energies of formation as a function of hydrogen gas pressure for various temperatures. The free energy of GNR formation changes systematically with both temperature and hydrogen pressure, mainly due to the change in the free energy of edge formation. These results allow us to construct a nucleation phase diagram (Fig. 3d) signifying that lower temperatures and higher hydrogen gas pressures favour GNR growth.

To verify the theoretical nucleation phase diagram, we carried out systematic growth experiments under different conditions. The obtained relative GNR and CNT populations are presented as pie charts on top of the phase diagram in Fig. 3d. Excellent agreement between our theoretical prediction and the experimental results is clearly seen. As an example, Figs. 3e-g present AFM images of post-growth surfaces under different hydrogen pressures. At a low pressure of $P_{H_2}$ = 0.01 atm, the samples consist of nearly pure CNTs (Fig. 3e). Increasing the hydrogen pressure to 0.25 atm, produces a mixture of CNTs and GNRs (Fig. 3f), whereas at a high hydrogen pressure of 0.5 atm pure GNR samples are obtained (Fig. 3g). This result is also supported by fully atomistic molecular dynamics (MD) simulations of the microscopic nucleation process, showing that only GNRs nucleate at sufficiently high hydrogen pressure (SI section 6b). We distinguish between CNTs and GNRs by AFM measurements of their height (> 1 nm for CNTs and 0.3 nm for GNRs), as shown in Fig. 3i. Notably, we find that when bent, the grown GNRs tend to produce sharper corners than CNTs, as may be expected from their relative bending stiffness.

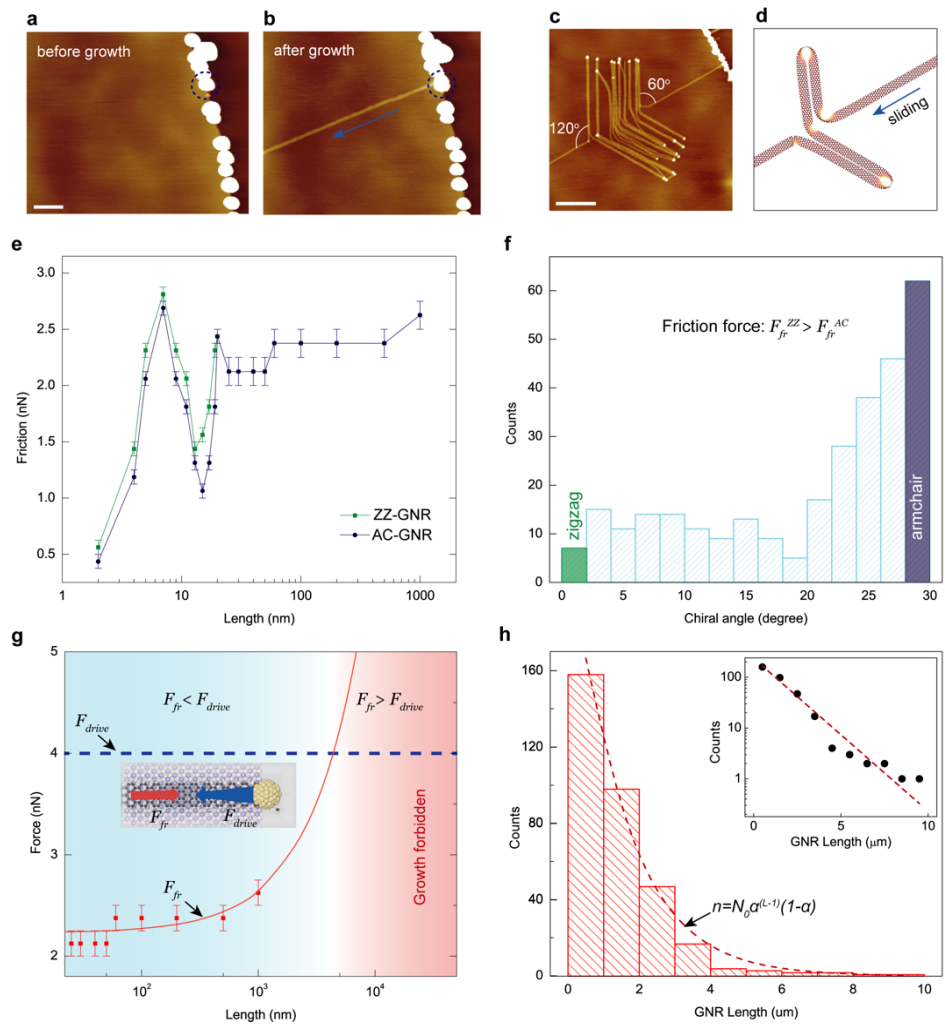

**Figure 4 van der Waals sliding growth of GNRs atop *h*-BN. a, b,** Experimental evidence of base-growth during GNR synthesis. Comparing the AFM images before and after growth, we conclude that the iron nanoparticle remains in place during the synthesis and the GNR grows away by sliding along the *h*-BN substrate. Scale bar, 100 nm. **c,** A peacock-like GNR structure, providing strong evidence of sliding motion of GNRs on *h*-BN along preferable lattice directions. Scale bar, 100 nm. **d,** Schematic of the structure of GNR in (c). **e,** Calculation of the friction force as a function of GNR length. For short ribbons, of the order of the moiré pattern dimensions, the friction varies strongly with the ribbon length. In this region, the friction experienced by AC-GNRs is systematically lower than that of ZZ-GNRs. **f,** Chirality distribution of GNRs grown on *h*-BN, where the amount of AC-GNRs is apparently larger than that of ZZ-GNRs. **g,** For long GNRs, when the friction force reaches the critical value, growth is terminated limiting the maximal GNR length obtainable. **h,** The length histogram of GNRs on *h*-BN matches the Schulz-Flory distribution (red dashed line). Inset: same data plotted on a semi-log plot.

Once nucleated, the GNRs continuously grow, pending the supply of carbon and hydrogen precursors. Figs. 4a and 4b establish that the nanoparticle catalysts migrate to the *h*-BN step edges during the heating-up stage and remain trapped there throughout the GNR growth process. This complies with a "base-growth" (rather than a "tip-growth") mechanism[19, 20, 32], where the catalysts remain in place while the growing GNR slides along the underlying *h*-BN substrate, suggesting that adhesion and friction may control their growth. Therefore, a detailed investigation of the friction experienced by the growing GNR, when it slides atop the *h*-BN substrate, is required. To that end, we performed fully atomistic MD simulations (see *Methods* section and SI Section 7) of the sliding motion of both ZZ- and AC-GNRs with different lengths on an *h*-BN surface. The results of these simulations, performed at the experimental growth temperature of ~800 °C, are presented in Fig. 4e. We find that for short GNRs with lengths comparable to the moiré period (~ 15 nm), the friction force varies with the ribbon length by a factor of ~ 2. Notably, the friction force found for short armchair ribbons $\left(F_{fr}^{AC}\right)$ is smaller than that of short zigzag ribbons $\left(F_{fr}^{ZZ}\right)$ by up to ~0.5 nN. Considering that the activation energy for the catalytic growth increases linearly with the resistive force[33], this translates to a formation rate ratio of (see *Methods* section): $e^{-(F_{fr}^{AC} - F_{fr}^{ZZ}) \cdot \delta / (k_B T)} \approx 120$, where $\delta = 1.42$ Å is the carbon-carbon bond length, $k_B$ is Boltzmann's constant, and $T = 1073\ K$ is the growth temperature. This suggests a strong preference towards AC-GNR growth along the AC direction of the *h*-BN substrate following the nucleation stage, in agreement with experimental observations (Fig. 4f).

Interestingly, GNR growth atop *h*-BN does not always proceed along a straight path. When the leading edge of the growing GNR reaches a surface obstacle, its growth direction may change via sideways slippage. Since the GNRs prefer to stack along certain lattice directions on the *h*-BN substrate, they tend to bend with angles of 60°, 120° and 180° (Fig. 3k and Fig. S17). This commensurability driven growth mechanism results in the peacock like structure appearing in Fig. 4c-d. A possible surface obstacle induced GNR folding mechanism is discussed in section 8 of the SI.

Finally, our simulations indicate a slow increase of friction with length for GNRs longer than 100 nm, suggesting a possible mechanism for GNR growth termination when the friction exceeds a critical value that can be calculated using mechanochemical considerations (see *Methods* section). The final length distribution of the fully grown GNRs plotted in Fig. 4h, matches the Schulz-Flory distribution[20] (SI section 9) and provides a maximum GNR length of ~10 μm (Fig. 4h).

We have therefore demonstrated the development of a method for scalable fabrication of free-standing, ultra-narrow, and micrometre-long GNRs on insulating *h*-BN substrates. The developed GNR growth method should also work for other ultra-flat substrates that form superlubric contacts with GNRs. Our findings provide exciting opportunities for studying electron transport properties of high-quality GNR devices. Specifically, electron-electron interaction induced correlation phenomena[34], such as Luttinger-liquid behaviour[35], Coulomb blockade[36], and Kondo effect, are expected to emerge due to the reduced electrostatic screening and enhanced electron-electron interactions in 1D structures. In addition, the 1D moiré potential can further flatten the electronic bands and reduce the kinetic energy of the charge carriers within the GNRs, making the moiré GNRs a promising platform to explore strongly correlated phenomena.

*Methods:*

**CVD growth of GNRs.**
*h*-BN flakes were mechanically exfoliated onto $SiO_2$/Si chips, which were then annealed in air at 600 ºC or exposed to hydrogen plasma at 300 ºC to remove all organic residual and contaminations. Catalytic nanoparticles (Fe) were deposited on the *h*-BN covered $SiO_2$/Si surfaces through thermal evaporation (evaporation rate: ~0.01 nm/s, base vacuum pressure: ~$1 \times 10^{-6}$ mbar). Then the chips were put into a tube furnace (Anhui BEQ Equipment Technology) and flushed with a mixture of hydrogen and argon for 3 min to remove other gas molecules. After that, the chips were gradually heated up to the growth temperature (600-900 ºC) under hydrogen and argon gas mixture at atmospheric pressure.

When growth temperature was reached, argon was replaced by methane to commence GNR growth. After a growth period of 5-60 min, the systems were cooled down to room temperature under a protective hydrogen and argon atmosphere. For GNRs growth on graphite substrates, the same Fe deposition and CVD growth recipe were used.

**Atomic force microscopy.**

A commercial AFM (Cypher S, Asylum Research, Oxford Instruments) was used to image the as-grown samples. GNRs were scanned in AC topography mode in air. AFM probes of AC200 and RTESPA-300 were typically used for the imaging. For high resolution AFM scanning in AC mode, PFQNE-AL and Arrow-UHFAuD probes were used. For lattice resolution imaging of *h*-BN, friction mode was used. Lattice information was obtained via fast Fourier transform.

**STM and STS measurements.**

Sample preparation and characterization were carried out using a commercial low-temperature Unisoku Joule-Thomson scanning probe microscopy under ultra-high vacuum conditions ($3\times10^{-10}$ mbar). GNRs were annealed at 200-400 ºC for 12 hours to remove any adsorbed air molecules under ultra-high vacuum. The samples were then transferred to a cryogenic scanner at 4.9 K for cooling. A lock-in amplifier (589 Hz, 10-30 mV modulation) was used to acquire $dI/dV$ spectra. The spectra were taken at 4.9 K unless otherwise stated.

**Calculations of free energy of formation and growth phase diagram.**

DFT calculations were performed using the Vienna *ab initio* Simulation Package (VASP)[37] with projected augmented wave (PAW) method[38]. The generalized gradient exchange-correlation functional approximation (GGA)[39] was utilized along with the D3 dispersion correction[40] to describe van der Waals interactions. The plane-wave cutoff energy was set to 600 eV. The Brillouin zone was sampled using Monkhorst-Pack k-mesh with a separation criterion of 0.02[41]. Criteria for energy and force convergence were $10^{-4}$ eV and $10^{-2}$ eV/Å, respectively.

To compare the thermodynamic stability of GNRs and CNTs at growth condition, we considered the influence of temperature and pressure to the formation of GNRs by adding the free energy change of GNRs from the hydrogen termination, which was estimated by the following equation:[42, 43]

$$\Delta G_f(\text{GNR}) = E_f(\text{GNR}) + \Delta F_{vib} - \frac{1}{2} N_H \times \mu_{H_2}, \quad (1)$$

where $E_f(\text{GNR})$, $\Delta F_{vib}$ $N_H$ and $\mu_{H_2}$ are the formation energy of GNRs, the vibrational entropy from the hydrogen termination at the GNR edge, the number of hydrogen atoms and the hydrogen chemical potential in gas phase. The first term was calculated using to the following equation:[44]

$$\Delta F_{\text{vib}} = \sum_\omega \hbar\omega \left(\frac{1}{2} + \frac{1}{e^{\beta\hbar\omega}-1}\right) - k_B T\left[\frac{\beta\hbar\omega}{e^{\beta\hbar\omega}-1} - \ln(1 - e^{-\beta\hbar\omega})\right], \quad (2)$$

where $\omega$ and $h$ are the phonon frequency and Plank's constant, and $\beta = (k_B T)^{-1}$. The second term was calculated using the following equation:[43]

$$\mu_{H_2} = H^0(T) - H^0(0\text{ K}) - TS^0(T) + k_B T \ln \frac{p_{H_2}}{p^0}, \quad (3)$$

where, the standard values of $H^0(T)$, $H^0(0\text{ K})$, and $S^0(T)$ were obtained from chemical tables[45], $p^0$ is the atmospheric pressure, and $p_{H_2}$ is the experimental hydrogen pressure. More details are presented in SI section 5a.

**MD simulation of the friction**.

The simulated model system consists of armchair and zigzag GNRs of fixed width (~2 nm) and different lengths in the range of 5-1000 nm sliding along the armchair or zigzag direction of a bilayer AA'-stacked *h*-BN substrate, respectively. During the simulation, the bottom layer of the substrate model was kept fixed. The GNRs' edges were passivated by hydrogen atoms[46] to avoid peripheral C–C bond reconstruction[47], that may influence friction. The intra-layer interactions within the GNRs and the *h*-BN substrate model were computed via the second generation of REBO potential[48] and the Tersoff potential[49], respectively. The interlayer interactions between the GNRs and the *h*-BN substrate were described via the registry-dependent ILP[50] with refined parametrization[51], which we implemented in LAMMPS[52]. High temperature simulations were performed adopting the following protocol. The initial configurations of the GNRs were generated via geometry optimization using the FIRE algorithm[53] implemented in LAMMPS[52] with a threshold force value of $10^{-6}$ eV/Å. This was followed by an equilibration step applied to bring the system to thermal equilibrium at 1073 K using a Langevin thermostat with a damping constant of 1 ps$^{-1}$ applied to the top layer of the $h$-BN substrate. Reducing the damping constant by an order of magnitude did not affect the results (see Section 6 of the SI). Sliding friction simulations with a fixed time step of 0.2 fs were then carried out by adding a constant pushing force on the three carbon atoms located at the trailing edge of the GNR (see Fig. S12 in Section 6 of the SI). To extract the static friction force, we gradually increased the pushing force with a finite step (0.0625-0.125 nN), until the GNR started sliding. The simulation ran at least 500,000 time steps (100 ps) for each pushing force. The static friction force was then defined as the average of the pushing force values right before and right after sliding commenced. Further details regarding the MD simulation at zero temperature and the effect of the value of the damping coefficients can be found in Section 6 of the SI.

**Relation between GNR growth velocity and its friction with the $h$-BN substrate.**
A quantitative description of the effect of the resistive friction between the GNR and the underlying $h$-BN substrate on its growth velocity can be deduced from mechanochemistry considerations. Here, the activation barrier for chemical reactions is assumed to vary linearly with an external force[33, 54]. For the reaction at hand, this suggests that the activation energy for the formation of a new GNR row at the catalyst surface atop the $h$-BN substrate reads as: $\Delta_f(F_{fr}) = \Delta_f^0 + F_{fr}(L_{\text{GNR}}) \cdot \delta$, where $\Delta_f^0$ is the activation barrier of the catalytic growth in the absence of the external friction force $(F_{fr})$ and $\delta$ is the length of a single GNR row. With this, the GNR growth velocity can be written as[55]:

$$v = \delta \left[ C_c \cdot k_f^0 \cdot e^{-\frac{F_{fr}(L_{\text{GNR}}) \cdot \delta}{k_B T}} - k_d \right], \qquad (4)$$

where, $C_c$ is the concentration of carbon precursors, $k_f^0 \propto e^{-\frac{\Delta_f^0}{k_B T}}$ is the rate of GNR row formation in the absence of the friction force, and $k_d$ is the GNR row decomposition rate, which is assumed to be independent of the external friction force. In our simulations, the friction force found for short armchair ribbons is smaller than that of short zigzag ribbons by up to ~0.5 nN, yielding a formation rate ratio of:

$$\frac{k_f^{AC}}{k_f^{ZZ}} = e^{-\frac{\left(F_{fr}^{AC} - F_{fr}^{ZZ}\right) \cdot \delta}{k_B T}} \approx 120, \qquad (5)$$

Eq. (4) also demonstrates that there is a critical friction force, $F_{fr}^c$, obtained when the term within the square brackets vanishes, above which GNR growth is inhibited:

$$F_{fr}^c = \frac{k_B T}{\delta} \ln\left(\frac{C_c \cdot k_f^0}{k_d}\right). \tag{6}$$

**Acknowledgments**

This work is supported by the National Key R&D Program of China (No. 2020YFA0309000), the National Natural Science Foundation of China (11774224, 12074244, 12102307, 11890673, 11890674, 11874258 and 12074247), by the Institute for Basic Science (IBS-R019-D1) of South Korea and by "Shuguang Program" supported by Shanghai Education Development Foundation and Shanghai Municipal Education Commission. W.O. acknowledges the starting-up fund of Wuhan University. M.U. acknowledges the financial support of the Israel Science Foundation, grant No. 1141/18 and the ISF-NSFC joint grant 3191/19. O.H. is grateful for the generous financial support of the Israel Science Foundation under grant no. 1586/17, Tel Aviv University Center for Nanoscience and Nanotechnology, and the Naomi Foundation for generous financial support via the 2017 Kadar Award. S.W. acknowledges support from Shanghai Municipal Science and Technology Qi Ming Xing Project (No. 20QA1405100), Fok Ying Tung Foundation for young researchers. K.W. and T.T. acknowledge support from the Elemental Strategy Initiative conducted by the MEXT, Japan, Grant Number JPMXP0112101001, JSPS KAKENHI (Grant Numbers 19H05790, 20H00354) and A3 Foresight by JSPS. S.W. and Z.S. acknowledges support from SJTU (21X010200846), and additional support from a Shanghai talent program. We would also like to acknowledge support from the Instrument Analysis Center of Shanghai Jiao Tong University (SJTU) for performing RISE and MIRA.

**Author contributions**

B.L. and Z.S. conceived the project. Z.S., S.W., F.D., M.U. and O.H. supervised the project. B.L., J.C. and S.L. grown samples with assistance from J.X. and T.W. B.L., J.C. and S.L. carried out AFM measurements. B.L. and J.C. preformed Raman measurements. C.L. and S.W. preformed low temperature STM measurements. L.Q. and F.D. calculated the phase diagram. W.O. designed the MD simulation setup and implemented the codes. W.O. and J.X. conducted MD simulations of GNR sliding. K.W. and T.T. grew *h*-BN single crystals. B.L., J.C., S.L., C.L., L.Q., W.O., J.X., T.W., M.U., O.H., F.D., S.W. and Z.S. analysed the data. B.L., J.C., S.L., W.O., M.U., O.H., F.D., S.W. and Z.S. wrote the paper with input from all authors. All authors discussed the results and edited the manuscript.